\documentclass[a4paper,12pt]{article}
\usepackage{amsmath}
\usepackage[utf8]{inputenc}
\usepackage{amssymb}
\usepackage{authblk}
\usepackage{graphicx,amssymb,amsfonts}

\newcommand\be{\begin{equation}}
\newcommand\ee{\end{equation}}
\newcommand\bea{\begin{eqnarray}}
\newcommand\eea{\end{eqnarray}}
\newcommand\del{\partial}
\newcommand\tld{\tilde}
\title{Topological dyonic Taub-Bolt/NUT-AdS:\\ Thermodynamics and first law}
\author[a,b]{Adel Awad \thanks{awad.adel@aucegypt.edu}}
\author[c]{Somaya Eissa\thanks{somaia@sci.cu.edu.eg}}

\affil[a]{\small \it Department of Physics, School of Sciences and
Engineering, American University in Cairo, P.O. Box 74, AUC Avenue
New Cairo, Cairo, Egypt} \affil[b]{\small \it Department of Physics,
Faculty of Science, Ain Shams University, Cairo 11566, Egypt}
\affil[c]{\small \it Department of Physics, Faculty of Science,
Cairo University, Giza 12613, Egypt}
\date{}
\begin{document}
\maketitle
\begin{abstract}
Motivated by the absence of Misner string in the Euclidean
Taub-Bolt/NUT solutions with flat horizons, we present a new
treatment for studying the thermodynamics of these spactimes. This
treatment is based on introducing a new charge, $N=\sigma \, n$ (where $n$ is the nut charge and $\sigma$ is some constant) and its conjugate thermodynamic
potential $\Phi_N$. Upon identifying one of the spatial coordinates,
the boundary of these solutions contains two annulus-like surfaces in
addition to the constant-r surface. For these solutions, we show that these annuli surfaces receive electric, magnetic and mass/energy fluxes, therefore, they have nontrivial contributions to these conserved charges. Calculating these conserved charges we find, $Q_e = Q^{\infty}_e-2N\Phi_m$, $Q_m =
Q^{\infty}_m+2N\Phi_e$ and $\mathfrak{M} =M-2N\Phi_N$, where $Q^{\infty}_e$, $Q^{\infty}_m$,
$M$ are electric charge, magnetic charge and mass in the $n=0$ case, while $\Phi_e$ and $\Phi_m$ are the electric and magnetic potentials. The
calculated thermodynamic quantities obey the first law of
thermodynamics while the entropy is the area of the horizon.
Furthermore, all these quantities obey Smarr's relation. We show the
consistency of these results through calculating the Hamiltonian and its variation which reproduces the first law.

\end{abstract}
\section{Introduction}\label{S1}
The last two decades have witnessed lots of interest in
Anti-de-Sitter (AdS) spacetimes and their thermodynamics as a result
of gauge/gravity dualities\cite{Maldacena1,GKP,Witten1}. These
dualities relate the physics of AdS spacetimes in the bulk at weak
coupling to that of conformal field theories on the boundary at
strong coupling. Such dualities have
 revealed how gravitational solutions in AdS encodes
important information about the conformal field theories on the
boundary in their semi-classical gravity. Black holes in
asymptotically AdS spacetimes have three types of horizon topology,
namely, spherical, hyperbolic and flat, in contrast to
asymptotically Minkowski black holes, which have only spherical
horizons. One of the interesting classes of AdS solutions which were
discussed in this context is the AdS-Taub-Bolt/NUT black holes in
four \cite{clifford98, surf99} and higher dimensions
\cite{adel+andrew}. These solutions
  were studied by serval authors \cite{mann,clifford98,surf99,mann-thermo,mann06} and have revealed unusual
  properties of their thermodynamics. For example, the entropy is not the area of
  the horizon as a result of Misner string and it is not
   always positive. Another unusual feature is that although
   we have added one more parameter to Schwarzschild-AdS solution we do not have an
   additional term in the first law, as in the case of Kerr solution when we add a rotation parameter
    "a". This is because of the known extra identification in the time direction $\beta=8\pi n$, which leaves Misner string invisible
    \cite{misner},
     where $n$ is the nut charge. This identification render the nut parameter related to, the horizon radius $r_0$, as a result, the nut charge can not be varied
     independently of the horizon radius $r_0$.


In this work we study thermodynamics and the first law of neutral
and dyonic Taub-Bolt/NUT-AdS solutions in four dimensions with flat
horizons, which are characterized by a NUT charge $n$ and have the
following form, \be ds^2=f(r)\,(dt+{\cal A})^2+{dr^2 \over
f(r)}+(r^2-n^2)\,d\Sigma_{B}^2. \ee where, ${\cal A}$ is the
one-form in the AdS-Taub-NUT/Bolt metric, which is related to the
Kahler form ${\cal F}$, through ${\cal F}=d{\cal A}$. Here we define
a new thermodynamical charge $N= -\int_{B} {\cal F} = \sigma n$,
where $\sigma$ is some constant\footnote{ e.g., in the case of
spherical horizon the integration is over $S^2$ and $\sigma=4\pi$.}.
The constant-r hypersurface of this space is a $U(1)$ fiber over the
base manifold {\bf B}. This solution has been generalized to higher dimensional Taub-Bolt and Taub-NUT solutions in
\cite{bais} for asymptotically locally flat spaces and in
\cite{page+pope,adel+andrew} for asymptotically locally de Sitter
and Anti-de-Sitter spaces.

Here we present a new treatment to study the thermodynamics of these
solutions in which the NUT charge, $n$, is directly related to the
new charge $N= -\int_{\Sigma} {\cal F} = \sigma n$. The treatment
presented here is motivated by the previous work of Hunter
\cite{hunter}, in which he introduces a charge $N=-1/4 \pi
\int_{S^2} {\cal F}$ and a nut potential $\psi_N$, which is similar
to the chemical potential $\Phi_N$ we define in this work. In our
treatment $N$ can vary independently of the horizon radius $r_0$,
since we do not have $\beta=8\pi n$, as a result of the absence of
Misner string in the Euclidean Taub-Bolt/NUT solution. As mentioned
above, the new charge $N$ has its own conjugate thermodynamic
potential, $\Phi_N$ and the $N \Phi_N$ terms is going to play an
important role in the thermodynamics of these solutions. Upon
calculating various thermodynamic quantities one finds that the
entropy is the area of the horizon and these quantities satisfy the
Smarr's relation and the first law.

Upon identifying one of the coordinates, the spatial boundary of
these cases consists of three surfaces, a constant-r surface and two
annulus-like surfaces extended from the horizon to infinity on the
top and the bottom of the cylinder. The two annulus-like surfaces in
$n=0$ case (e.g., Schwarzschild-AdS and Reissner-Nordstorm-AdS), do
not receive any fluxes of conserved quantities such as mass,
electric and magnetic charges, therefore, one can consider the
identification of the other spatial direction obtaining a torus. But
as we will see here for $n\neq 0$ these annulus-like surfaces
receive their own fluxes which are going to contribute in the
surface integrals of various conserved charges. In fact, we are
going to show that these surfaces have nontrivial contributions to
all conserved quantities, such as, electric charge, magnetic charge
and the mass of the solution.

The thermodynamics of these solutions has been studied by Mann et
al. in \cite{mann06}. In their thermodynamics treatment the mass of
the solution is the usual mass $M=\sigma\, m$, which is proportional
to the mass parameter $m$, but the entropy is not the area of the
horizon, since it needs to satisfy the Gibbs-Duhem relation,
$I=\beta\,M-S$. In section 3, we show that a careful calculation of
the mass using Komar integral reveals an additional contribution
coming from the annulus integrals of the cylinder which add a
$-2N\Phi_N$ term to the mass. Although, the quantities in
\cite{mann06} satisfy Gibbs-Duhem relation, they do not satisfy the
first law unless the outer radius of the horizon $r_{0}$ is related
to the nut charge. This relation is very similar to the one found in
the spherical horizon case where one imposes the extra periodicity
condition on Euclidean time $\beta=8\, \pi n$ to remove the Misner
string singularity. In the spherical case this relation is important
to remove such a singularity but for the flat horizon case we have
no such a singularity and no Misner string, therefore, it is not
clear how can we justify such a relation. Another important point is
that in the absence of Misner strings, we expect the entropy of the
solution to receive contributions solely from the area of bolts
\cite{hunter+hawking}, since the known expression of entropy in this
case is $S= {1 \over 4} ({\cal A}_{bolt}+{\cal A}_{MS})-\beta \,
H_{MS}$, i.e., the entropy is the area of the horizon, therefore,
this analysis is not consistent with these known results.

Recently, the authors in \cite{mann19} have considered a similar
treatment to the one we present here, but for Lorentzian spherically
symmetric Taub-Bolt solutions, rather than an Euclidian Taub-Bolt
solution with flat horizon. In their treatment the new charge "N" is
not directly proportional to $n$ but a function of $n$ and horizon
radius $r_0$ that changes from a solution to another \cite{mann06,
BGK}. A second difference is that our mass calculation gives an
additional contributions from the annulus-like surfaces which gives
a total mass, ${\mathfrak{M}}=M-2\,N \Phi_N$, where $M={\sigma}
\left(m-m_n\right)={\sigma} (m+{4n^3/l^2})$ and we have used the NUT
space as a reference space. This leads to the first law,
$d\mathfrak{M}=\,T\, dS-N \, d\Phi_N$, which is different from the
one obtained in \cite{mann19}. A third difference is that the
chemical potential $\psi_N$, associated with their New charge "N"
was shown to be proportional to Misner string temperature
\cite{BGHK}. But since we deal with multi-temperature system, at
equilibrium they should match, which is equivalent to identifying
the time direction with periodicity $\beta=8\pi n$. Although our
treatment and the one in \cite{mann19} have the same general idea of
introducing a pair of thermodynamical variables $N-\Phi_N$, which is
a natural consequence of having independent $r_0$ and $n$, the two
approaches are different in their mass and their first law looks
different as we will see below. While preparing this work we found the appearance of another related paper \cite{ww} which introduces again the same general idea of adding extra thermodynamic pairs, but this time it adds two new charges $n$ and $m\,n$ to study the thermodynamics of Lorentzian Taub-Bolt and NUT solutions with spherical horizons. Although this work studies a different class of Taub-Bolt/NUT solutions with different signature which includes electrically charged Taub-Bolt solutions it did not include the dyonic solution we studied here.

This work is organized as follows: In section $2$ we discuss the
thermodynamical consequences of having the new thermodynamic
quantity $N$ in analogy with electric and magnetic charges. In
section $3$ we calculate the total mass of the spacetime given the
additional contributions of the annulus-like surfaces, the entropy
as the area of the horizon and show that the first law
and Smarr's formula are satisfied when we include $N$ and its
chemical potential $\Phi_N$. In section $4$ and $5$ we calculate all
relevant thermodynamic quantities, then show that they do satisfy
the first law and Smarr's formula when we include $N$ and its
chemical potential $\Phi_N$ as well as the additional contribution
of the magnetic charge and total mass. In section $6$, we calculate
the Hamiltonian and its variation to reproduced the first law that
we obtained in the previous section, which confirms its form. At the
last section, $7$ we conclude our work.

\section{Action and Thermodynamic Ensembles}

In this section we discuss the thermodynamic consequences of
introducing the new charge "N" and its chemical potential for the
neutral and dyonic AdS-Taub-Bolt solutions and argue for an alternative
treatment in which the entropy is the area of the horizon and the
first law is satisfied without the need for the extra condition
relating the radius of the horizon to the NUT charge. The action of
Einstein-Maxwell theory for asymptotically AdS spacetime ${\cal M}$
with boundary $\del{\cal M}$ is given by
\begin{equation} \label{Ibulk}
    I_G\,=-\frac{1}{16 \pi G}\,\int d^4 x \, \sqrt{-g}\, (R-2\Lambda-F^2) -\frac{1}{8\pi G} \int_{\del {\cal M}} d^3x
    \, \sqrt{h}\,K,
\end{equation}
where $\Lambda=-\frac{3}{l^2}$ is the cosmological constant, $A_{\mu}$ is the gauge potential
and $F_{\mu\nu}=\del_{\mu}A_{\nu}-\del_{\nu}A_{\mu}$ is its field
strength. The first two terms represent the Einstein-Hilbert action
with negative cosmological constant and the electromagnetic
contribution to the action. The final term is Gibbons-Hawking
boundary term. Here, $h_{ab}$ is the boundary metric and $K$ is the
trace of the extrinsic curvature $K^{ab}$ on the boundary. Varying the action with respect to the metric $g_{\mu \nu}$ and the gauge potential $A_{\mu}$ we get the following field equations

\begin{equation}
G_{\mu \nu}\,+\,\Lambda\,g_{\mu \nu}\,=\,2\,T_{\mu \nu}
\end{equation}
\begin{equation}
    \partial _{\mu}(\sqrt{-g}\,F^{\mu \nu})\,=\,0
\end{equation}
where $G_{\mu \nu}$ is Einstein Tensor, $T_{\mu \nu}$ is the stress tensor, which is given by

\begin{equation}
    T_{\mu \nu}\,=\,F_{\alpha \mu}\,F^{\alpha}_{\nu}-\frac{1}{4}g_{\mu \nu}\,F^2
\end{equation}
 \newline
Trying to calculate the above gravitational action for the Taub-Bolt
solution on shell, one encounters a finite number of divergent terms
arising from integrating over the infinite AdS volume. These
divergent terms can be canceled through the addition of certain
local surface counterterm \cite{bala99,surf99} or through the the
background subtraction technique where an action of a background
space (such that of a AdS space) is subtracted from the
 action of the spacetime under investigation. Here we are going to use the AdS-Taub-NUT space as our background spacetime.
 In our case here the boundary metric
is not a single hypersurface but a two-dimensional cylinder upon
identifying one of the dimensions. Therefore, one
 expects having additional attributions from the two annulus-like surfaces of
the cylinder in the action as well as the conserved charges of the
solution. From various studies of charged and neutral AdS black
holes with flat horizons in literature, one can see that these
surfaces have vanishing contributions, when $n=0$. But for AdS
solutions with nonvanishing NUT charge, $n$, these surfaces have
nontrivial contribution to conserved quantities, such as, electric
charge, magnetic charge and the mass of the solutions, as we will
see in the coming sections.

Electrically charged black holes differ from the magnetically
charged ones in their boundary conditions \cite{CPW,hawking+ross}.
The boundary condition on the Euclidian action of magnetic black
holes fixes the magnetic charge, $Q_m$, therefore, the partition
function $Z=Z(T,Q_m)$, but for the electric black holes it fixes the
electric potential, $\Phi_e$ (chemical potential), therefore, the
partition function $Z=Z(T,\Phi_e)$. As a result, it is natural to
consider the canonical ensemble for the magnetic case and the grand
canonical ensemble for the electric case. This leads to a mixed
ensemble in the case of dyonic black hole with the partition
function $Z(T,Q_m,\Phi_e)$. In the case of AdS-dyonic black holes
with spherical horizons, one can consider the canonical ensemble with fixed charges upon adding a surface term to the action, which reads
\be \tilde{I}=I-{1 \over 4\pi G}\int_{\del{\cal M}} d^3x\sqrt{h}\,
n_{a}\,F^{ab}\,A_a. \ee It provides the action with the needed
Legendre transformation to replace its dependence on $\Phi_e$ with a
dependence on $Q_e$.

It is worth mentioning here that from a thermodynamic perspective,
the NUT charge is quite similar to the magnetic charge since both
charges are expressed as integrals of some field strength over a
closed surface at the boundary, therefore, these charges are fixed
upon fixing the boundary metric. As a result, we evaluate the
partition functions in a definite charge sector, or $Z(T,N)$, where
the action is related to the Gibbs energy. Then, \be \left({\del G
\over \del N}\right)_{T} =\Phi_N, \ee defines the chemical potential
for $N$ \footnote{In \cite{hunter} Hunter argued for the existence
of a similar term to $n\, \Phi_n$ in the action of Taub-NUT space
with spherical horizon.}. This leads to \be d G = -S\,dT +\Phi_N d
N,\ee as we are going to see in the coming sections.

\section{Taub-Bolt thermodynamics with flat horizon revisited}

As was mention in the previous sections our thermodynamic treatment
of Taub-Bolt/NUT-AdS solutions assumes that $N$ is a new charge
which can be varied independently from the horizon radious. It is
constructive to discuss the neutral Taub-Bolt/NUT case before we
move to the Taub-Bolt/NUT case with electric and magnetic charges.
The Taub-Bolt/NUT spacetime is given by the following metric
\begin{equation}
    dS^2=f(r)\left(dt+{2\,n\,x \over l}d\phi\right)^2+\frac{dr^2}{f(r)}+\left(\frac{r^2-n^2}{l^2}\right)(dx^2+l^2\,d\phi^2)
\end{equation}
where
\begin{equation}
    f(r)=\frac{r^4-6n^2r^2-3n^4-2ml^2r}{l^2(r^2-n^2)}
\end{equation}
Here, $r$ is a radial coordinate, $x\in [-L/2,L/2]$ and $\phi \in
[0,2\pi]$. Total surface area at constant r is $\int l\, d\phi \,
dx=2\pi\,l\,L=\sigma\,l^2$. Notice that the spatial boundary is a
two-dimensional cylinder. The horizon radius $r_0$ is defined as
$f(r_0)=0$.

One of the distinguishing features of this solutions is the absence of Misner string (for example see appendix C \cite{mann06} ), which means that we do not have the additional condition $f'(r_0)={1 \over 2n}$, as a result, the temperature is $T=f'(r_{0})/4\pi$ \cite{clifford98,surf99}. Since there are no contributions from Misner string, the entropy is just the area of the horizon (i.e., the Bolt contribution), in contrast with the spherical horizon case. The condition for a nut is that the mass parameter $m=m_n=-4n^3/l^2$ \cite{clifford98,surf99}, which leads to a zero-dimensional fixed point rather than a two-dimensional fixed point which we call a bolt.

The temperature at the horizon is proportional to surface gravity of the black hole and is given by,
\begin{equation} \label{T}
    T=\frac{f^{\prime}(r)}{4\,\pi}=3\frac{(r_0^2-n^2)}{4 \pi l^2 r_0}.
\end{equation}
Calculating the on-shell action using the background method using
NUT spacetime as a background space, one gets the action \be
I=\frac{\sigma\,\beta}{2}\,\left( m+\frac{r_0(3n^2-r_0^2)+2n^3}{l^2}
\right). \ee

Our analysis is based on the on-shell gravitational action $I(T,N)$
which we are going to use to calculate the entropy and the chemical
potential $\Phi_N$ of the conserved charge $N=\sigma\,n$. The change
in Gibbs energy $G(T,N)=I/\beta$ is \be dG=-S\,dT+\,\Phi_N \,dN, \ee
where \be \left({\del G \over \del T} \right)_N=-S, \, \hspace{0.6
in} \left({\del G \over \del N} \right)_T=\, \Phi_N.
\label{gibbsv}\ee

Using equation (\ref{gibbsv}), or varying the action with respect to
$\beta=1/T$, one can obtain the entropy from the action \be S=\beta
\del_{\beta}I-I, \ee which is the area of horizon, in contrast with the entropy calculated in \cite{mann06},
\begin{equation}\label{S}
    S=\frac{A_H}{4\,G}=\sigma\,{\pi (r_0^2-n^2)}.
\end{equation}
\begin{center} \centerline{ \includegraphics[angle=0,width=120mm]{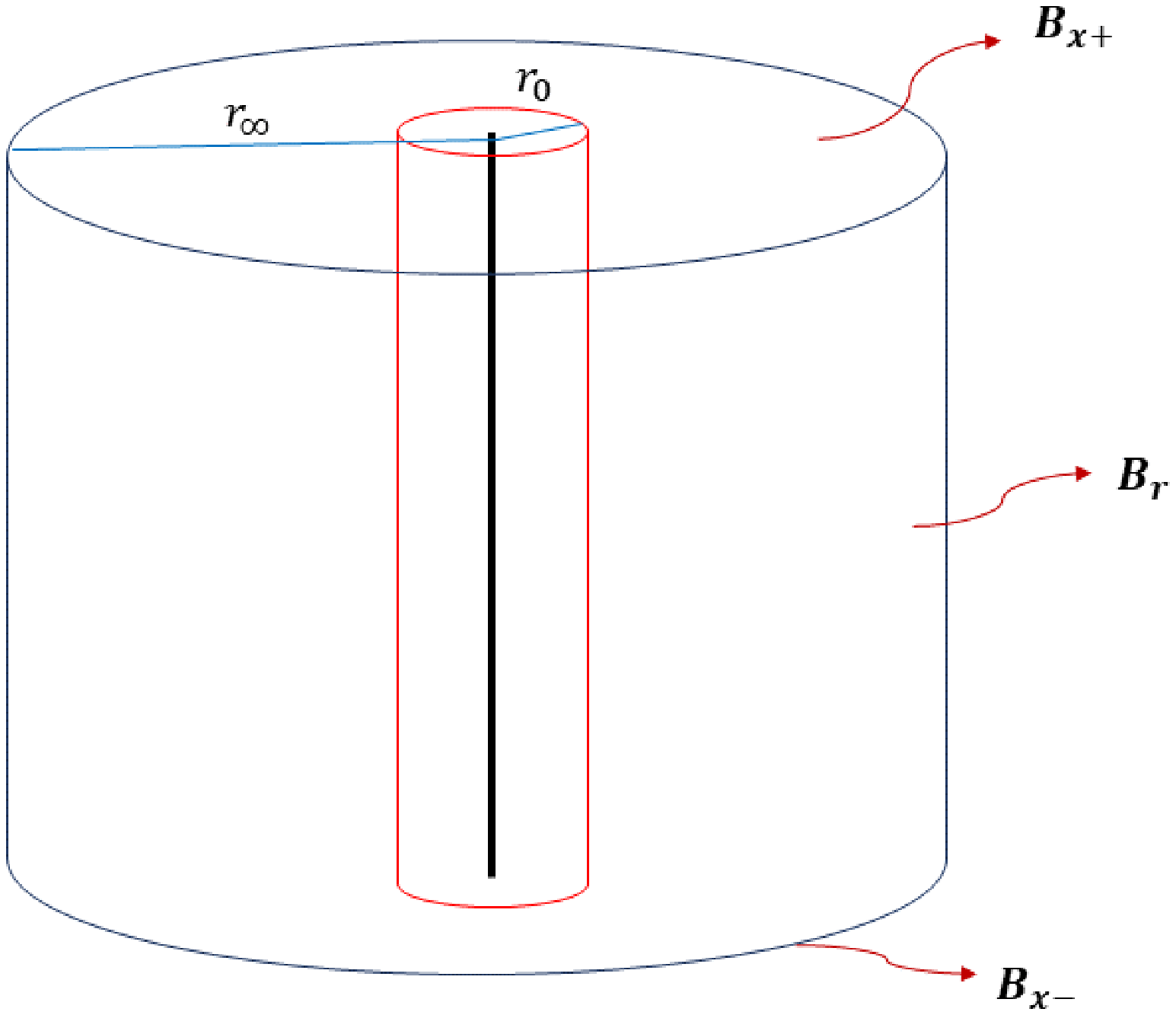}} {\footnotesize Figure 1: The boundary manifold at a constant time form a closed two dimensional paces with three
surfaces that $B_{r}$, $B_{x+}$ and $B_{x-}$ } \label{fig-1}
\end{center} The total mass ${\mathfrak{M}}$ can be obtained through
a careful calculation of Komar mass integral which have three
contributions, since the two-dimensional boundary is a cylinder-like
surface with a top and a bottom annuli, $B_{x+}$ and $B_{x-}$. These
are constant-x surfaces, in addition to $B_{r}$ the constant-r
surface at large $r$ as shown in Figure-1. \be
\mathfrak{M}=\int_{B}d\sigma^{\mu \nu}\,\zeta_{\mu;\nu}=\int_{B_{r}}d\sigma_r^{\mu \nu}\,\zeta_{\mu;\nu}
+\int_{B_{x+}}d\sigma_{x_{+}}^{\mu \nu}\,\zeta_{\mu;\nu}
+\int_{B_{x-}}d\sigma_{x_{-}}^{\mu \nu}\, \zeta_{\mu;\nu},\ee
where, $\zeta=\partial_t$ is a time-like Killing vector. The annulus integrals give vanishing contributions for $n=0$ case, and the
only nonvanishing contribution is coming from the constant-r surface
at infinity. But for $n\neq 0$ case, these annulus integrals leads to
additional terms in the total mass, namely, $2N\,\Phi_N$, leading to
${\mathfrak{M}}=M-2N \Phi_N$, where $M={\sigma}
\left(m-m_n\right)={\sigma}(m+{4n^3/l^2})$ and we have used Taub-NUT space
as a background space. Also, this mass might be obtainable from the
counterterm method, keeping in mind that we have three surfaces that
form the boundary at infinity, in this calculation one obtain an
additional, $2N\Phi_N$ term from the annulus integrals, but there is a
linear divergent term $~ n^2\, r$, which we could not remove using
the known counterterms.

Another way of calculating the total mass of the spacetime is through 
the relation \be {\mathfrak{M}}=\del_{\beta}I=(I+S)\,T-N\,\Phi_N,\ee
which is given by the following expression \be \mathfrak{M}=\sigma
{\left[ 2n^2+(r_0+n)^2\right]\,(r_0-n)^2 \over 2\,l^2\,r_0}, \ee we
have also found that \be \Phi_n=\left({\del \mathfrak{M} \over \del
n}\right)_{S}=-\frac{3\,n \,\sigma\,(r_0-n)^2}{2\,r_0\l^2}.\ee
Varying this mass with respect to the entropy, one finds \be
\left({\del \mathfrak{M} \over \del S}\right)_{N}=\, T.\ee One can
check that the first law is satisfied or \be d\mathfrak{M}=\,T\,
dS-N \, d\Phi_N. \ee Notice that the mass, $\mathfrak{M}$ is not the
internal energy but its Legendre transformation, since the internal
energy depends on extensive quantities such as conserved charges
rather their chemical potentials. This is very similar to
identifying the mass with enthalpy when we introduce the
cosmological constant as pressure in extended black hole
thermodynamics. Here we define another type of energy $M_H$,  \be
\mathfrak{M}=M_H-N\,\Phi_N, \ee where \be
dM_H=d\mathfrak{M}+d(N\,\Phi_N)=\,T\, dS+\Phi_N\, dN. \ee $M_H$
depends on the extensive quantities $S$ and $N$, which behaves more
like an internal energy for these variables and it is the Legendre
transform of the total mass. We will see that upon adding pressure
(allow cosmological constant to vary), $M_H$ is not the internal
energy but the enthalpy of the spacetime. Let us stop here for a
moment to discuss the meaning of the negative chemical potential
$\Phi_N<0$. One notices that in order to increase $N\rightarrow
N+\delta N$, the spacetime must do some work, or spent some energy
for this change to take place. This energy is nonvanishing as long
as $r_0>n$, but as $r_0\rightarrow n$, this amount is going to be
vanishingly small.

An important issue here which needs some discussion is the effects
of these annulus integrals on the action calculation. These
integrals might modify the action through adding new surface terms
to the Gibbons-Hawking surface term. In the previous Komar mass
calculation, although the extrinsic curvature $K_{ij}$ of the
constant-x hypersurface is not vanishing in the $n\neq 0$ case
(which contribute to the mass), its trace is vanishing, as a result,
the action expression is not changed. This explains why we keep
using the action with the Gibbons-Hawking term at constant-r
surface.

If we allow the cosmological constant $\Lambda=-{3 \over l^2}$ to
vary\footnote{To describe a dynamically varying cosmological constant in gravitational theories one might introduce a sourced totally anti-symmetric tensor field couples to  gravity \cite{brown1987dynamical,brown1988neutralization,duff1980quantum,nicolai1980}. In the classical equations of such a system the anti-symmetric tensor gives rise to a term that acts as cosmological constant. In this work we consider the cosmological constant as a thermodynamic variable following the  phenomenological approach suggested and developed by the authors in \cite{enth-1,enth-2,enth-3,enth-4} which interprets it as a pressure since its conjugate variable is the volume. This approach is sometimes called extended thermodynamics.}, according to extended
thermodynamics\cite{enth-1,enth-2,enth-3,enth-4}, we get a pressure
$P=-{\Lambda \over 2}$, the thermodynamic volume is given by \be V=
\left({\del G \over \del P}\right)_{S,N}={\sigma \over
3}\,(2n+r_0)(r_0-n)^2,\ee which is always positive for $r_0>n$.
Notice that the volume here is different from the one discussed in
\cite{clifford14-1,clifford14-2} or \cite{mann19} since our action
is $I=I_b-I_n$, i.e., our reference spacetime is the Taub-NUT
instead of AdS. It is relevant here to see the variation of
different thermodynamics quantities, \be d
M_H=TdS+\Phi_N\,dN+VdP,\ee \be d
{\mathfrak{M}}=TdS-N\,d\Phi_N+VdP,\ee where, \be \left({\del M_H
\over \del N}\right)_{S,P}=\Phi_N, \hspace{1.0 in} \left({\del M_H
\over \del P}\right)_{S,N}=V, \ee \be \left({\del \mathfrak{M} \over
\del \Phi_N}\right)_{S,P}=-N, \hspace{1.0 in} \left({\del
\mathfrak{M} \over \del P}\right)_{S,\Phi_N}=V. \ee Testing the
above expressions through Smarr formula, one find that the above
quantities satisfy \be {\mathfrak{M}}=2\, S\,T-2\,P\, V,\ee
 which is nontrivial test for the consistency of this treatment and in the case of vanishing $\Lambda$ it reduces to the same expression given by Hunter \cite{hunter}.
Notice here that total mass is neither the internal energy, $U$, nor
enthalpy but \be {\mathfrak{M}}=U+PV-N\Phi_N=H-N\Phi_N,\ee or the
Legendre transform of the enthalpy, $H=M_H$ since it depends on
$\Phi_N$ rather than $N$.
\section{Dyonic-Taub-(Bolt/NUT) Black Holes with Flat Horizon}

The dyonic-Taub-Bolt/NUT solution has the following form for the
metric
\begin{equation}
    dS^2=f(r)\left(dt+{2\,n\,x \over l}d\phi\right)^2+\frac{dr^2}{f(r)}+\left(\frac{r^2-n^2}{l^2}\right)(dx^2+l^2\,d\phi^2)
\end{equation}
where
\begin{equation}
    f(r)\,=\,\frac{r^4-6\,n^2\,r^2+l^2\,p^2-3\,n^4-l^2\,q^2-2\,m\,l^2\,r}{l^2\,(r^2-n^2)},
\end{equation}
and the nonvanishing components of the gauge potential are
\begin{equation}
    A_t\,=\,\frac{n\,p+V_e(\,n^2-r^2\,)-q\,r}{n^2-r^2}
\end{equation}
\begin{equation}
    A_{\phi}\,=\,\frac{p\,(n^2+r^2)-2\,n\,r\,q}{l^2\,(r^2-n^2)}\,x
\end{equation}

Thermodynamics imposes some regularity conditions on the gauge
potential $A_\mu$ at the horizon which are different for the Bolt
and NUT cases. For the Bolt case, it requires the following
relation,
\begin{equation} q_b\,=\,\frac{n\,p_b+V_e\,(n^2-r_{0}^2)}{r_{0}},
\end{equation}
For the NUT case, the $A_\mu$ regularity condition requires, \be
p_n=-2\,n\,V_e, \hspace{0.6 in} q_n=-2\,n\,V_e, \ee where, these
conditions leads to $f(r=n)=0$, for $m=m_n=-4\,n^3/ l^2$.


The temperature at horizon is given by,
\begin{equation} \label{T}
    T\,=\,\frac{f^{\prime}(r)}{4\,\pi}\,=\,\frac{1}{4\,\pi\,l^2\,r_{0}^3}\,\left[l^2\,V_e^2\,(r_{0}^2-n^2)\,+\,3\,r_{0}^4-l^2\,p\,(p+2\,n\,pV_e)-3\,r_{0}^2\,n^2\right].
\end{equation}
Here we have electric and magnetic potentials, $\Phi_e$ and $\Phi_m$, which can be calculated as follows
\begin{equation} \Phi_e\,=\,A_t\bigg\rvert_{\infty}\,-\,A_t\bigg\rvert_{r_{0}}\,=\,V_e
\end{equation}
\begin{equation}
    \Phi_m\,=\,\Tilde{A_t}\bigg\rvert_{\infty}\,-\,\Tilde{A_t}\bigg\rvert_{r_{0}}\,=\,\frac{p\,+\,n\,V_e}{r_{0}},
\end{equation}
where, $\Tilde{A_t}$ is defined as $\Tilde{F}=d\Tilde{A}$ and
$\Tilde{F}$ is the dual field strength.

Magnetic and electric charges are defined as the surface integral of
the electromagnetic tensor $F$ and its dual $\Tilde{F}$
 \be
 { Q}_m\,=-\int_{B_\infty}\,F, \hspace{0.7 in}  { Q}_e\,=-\,\int_{B_\infty}\,\Tilde{F}. \ee
Here it is important to notice the nontrivial contributions coming
from the two annuli $B_{x\pm}$, which are going to add new
contributions to the the electric and magnetic charges at the
boundary compared to their values when $n$ is vanishing.
\begin{equation} { Q}_m\,=-\int_{B_\infty}\,F\,=-\,\int_{B_r}\,F_{xy}\,dx\,dy\,-\,\int_{B_{x\pm}}\,F_{yr}\,dy\,dr,
 \end{equation}
Then, the total magnetic charge is given by
\begin{equation}\label{finalP}
    { Q}_{m}\,=\,{\sigma}\,(\,p\,+\,2\,n\,V_e\,).
\end{equation}
The electric charge $Q_e$ is defined as
\begin{equation}
    { Q}_e\,=-\,\,\int_{B_\infty}\,\Tilde{F}\,=-\,\int_{B_r}\,\Tilde{F}_{xy}\,dx\,dy\,-\,\int_{B_x}\,\Tilde{F}_{yr}\,dy\,dr.
\end{equation}
Following the same analysis, one gets
\begin{equation} \label{Q}
    {
    Q}_e\,=-\,\frac{\sigma}{r_{0}}\,({n\,p\,+\,V_e\,n^2\,+\,V_e\,r_{0}^2}).
\end{equation}
The total conserved electric charge $Q_e$, can be written as, \be {
Q}_e= Q^{\infty}_e-2\,N\,\Phi_m,\ee where  $Q^{\infty}_e=\sigma\,q$,
is the total charge when we set $n=0$. This electric charge is going
to play a role in thermodynamics rather than ${ Q}_e$. Also, the
total conserved magnetic charge $Q_m$, can be written as, \be {
Q}_m=Q^{\infty}_m+2\,N\,\Phi_e, \ee where,  $Q^{\infty}_m=\sigma\,p$
is the total magnetic charge when we set $n=0$. The electric and
magnetic potentials can be written in the following form \be \Phi_e=
-{Q_e+N\,\Phi_m \over \sigma \, r_0}, \hspace{0.4 in} \Phi_m=
{Q_m-N\,\Phi_e \over \sigma \, r_0}.\ee

\section{Thermodynamics and the first law}
Calculating the on-shell action using the background method and
choosing Taub-NUT as the reference space, one gets
\begin{equation}
I\,=\,\frac{\sigma\,\beta}{2}\, \left[\,m-q\,V_e+(p+n\,V_e)(p\,+2\,n\,V_e\,)/{r_{0}}+{r_{0}}\,(\,3\,n^2-r_{0}^2)/{l^2}\,+2n^3/l^2\right].\label{I}
\end{equation}
Again, varying the action with respect to $\beta$, one can calculate the entropy from the action \be S=\beta \del_{\beta}I-I, \ee
\begin{equation}\label{S}
     S\,=\,\frac{A_H}{4\,G}\,=\,\pi\,A_H\,=\sigma\,{\pi\,({r_0}^2-n^2)},
\end{equation}
which is the area of horizon. Let us check the thermodynamic
quantities of these solutions, at the same time, allow for a varying
cosmological constant, or pressure, $P={3 \over 2\,l^2}$. The Gibbs energy
$G(T,N,\Phi_e,Q_m,P)$ is nothing but $I/\beta$. Notice that the volume is same as
the noncharged solution, $V={\sigma \over 3}\,(2n+r_0)(r_0-n)^2$.
One can check the consistency of
thermodynamics through the variation of $G$
\be \left({\del G \over \del \Phi_e}
\right)_{T,\Phi_e,Q_m,P} =-Q^{\infty}_e, \hspace{0.5 in} \left({\del
G \over \del T} \right)_{N,\Phi_e,Q_m,P} =-S, \ee \be \left({\del G
\over \del Q_m} \right)_{T,N,\Phi_e,P} =\,\Phi_m, \hspace{0.3 in}
 \left({\del G \over \del P}\right)_{T,N,\Phi_e,Q_m} =V. \ee
In addition, one can calculate $\Phi_N$, as \be \left({\del G \over \del N}\right)_{T,\Phi_e,Q_m,P} =\Phi_N\, \ee
\be \Phi_N=-\, {1\, \over 2\,r_0^3}\,\left(2\,p_m\,(\,n\,p_m+\,V_e\,(r_0^2-n^2))\,+n\,{V_e}^2\,(\,n^2-3\,r_0^2)\,+{3\,n\,(n-r_0)^2\,r_0^2 \over l^2}\right), \ee where, $p_m=p-2\,n\,V_e$. These results are consistent with the change in $G$.
\begin{equation}\label{FL}
    d
    {G}\,=-S\,dT\,-\,Q^{\infty}_e\,d\Phi_e\,+\Phi_m\,d\,Q_m\,+\,\Phi_N\,dN+V\,dP,
\end{equation}
Defining the total energy/mass of the spacetime as
\be {\mathfrak{M}}=\del_{\beta}I+Q^{\infty}_e\,\Phi_e, \label{tmass}\ee
one gets the following expression
\be
   {\mathfrak{M}}\, = \,{\sigma \over 2\,r_0\,}\left[ {[2n^2+(r_0+n)^2 ](r_0-n)^2 \over l^2}+ {\left[ (r_0^4+4\,n^2\,r_0^2-n^4)\,(q^2-p_m^2)-8\,n^3\,r_0\,q\,p_m\right] \over (r_0^2+n^2)^2} \right].\ee
Similar to the neutral case the mass of the solution can be written
as \be {\mathfrak{M}}\,=M-2\,N\,\Phi_N \ee The first law of
thermodynamics for the charged solution takes the form
\begin{equation}\label{FL}
    d {\mathfrak{M}}\,=\,T\,dS\,+\,\Phi_e\,d\,Q^{\infty}_e\,+\Phi_m\,d\,Q_m\,-N\, d\Phi_N+V\,dP.
\end{equation}
The first law can be checked through the following relations
\bea && \left({\del {\mathfrak{M}} \over \del Q_m}\right)_{S,Q^{\infty},\Phi_N,P} =\Phi_m ,\hspace{0.5 in}\left({\del {\mathfrak{M}} \over \del S} \right)_{\Phi_N,Q^{\infty}_e,Q_m,P}=T, \hspace{0.3 in} \left({\del {\mathfrak{M}} \over \del \Phi_N}\right)_{S,Q^{\infty},Q_m,P} =-N , \hspace{0.3 in} \nonumber\\
  && \left({\del {\mathfrak{M}} \over \del Q^{\infty}_e} \right)_{S,\Phi_N,Q_m,P} =\Phi_e, \hspace{0.3 in} \left({\del {\mathfrak{M}} \over \del P}\right)_{S,\Phi_N,Q^{\infty}_e,Q_m} =V.\eea
The change in enthalpy, \be
H=M_H=M-N\,\Phi_N={\mathfrak{M}}+N\,\Phi_N \ee takes the following
form \be dH=TdS+\Phi_N\,dN+\Phi_e\,d\,Q^{\infty}_e\,+\Phi_m\,d\,Q_m+V\,dP.
\ee Again, testing these expressions through Smarr' formula, one
find that the above quantities satisfy
\begin{equation} {\mathfrak{M}}=2\, S\,T-2\,P\, V\,+\,Q^{\infty}_e\,\Phi_e\,+\,Q_m\,\Phi_m,\end{equation}
 which is a sign for the consistency of this approach. If we substitute with $\mathfrak{M}=M-2N\Phi_N$,
 our Smarr's relation is identical to the one obtained recently in \cite{smarr-TN} for the spherical Lorentzian Taub-Bolt
 case. It is interesting to see here that the electric charge that enter the first law is $Q^{\infty}_e$ but the magnetic charge is the full charge $Q_m$. In order to confirm these results formally, we are going to express the mass variation using Hamiltonian calculations. This is the task of the coming section.

\section{Hamiltonian and First Law} Following the Hamiltonian calculation in \cite{horowitz} and its variation one can confirm the previous results. To calculate the Hamiltonian we analytically continue the dyonic solution to get its Lorentzian version. Let us take,
\be t=i\, \tau, \hspace{0.5 in},
n=i\, \tld{n},\hspace{0.5 in},q=i\,\tld{q},\hspace{0.5 in} V_e=-i\,
\tld{V}_e,\hspace{0.5 in}, p=\tld{p}.\ee Our solution becomes
\begin{equation}
    dS^2=-f(r)\left(d\tau+{2\,\tld{n}\,x \over l}d\phi\right)^2+\frac{dr^2}{f(r)}+\left(\frac{r^2+\tld{n}^2}{l^2}\right)(dx^2+l^2\,d\phi^2),
\end{equation}
where,
\begin{equation}
    f(r)\,=\,\frac{r^4+6\,\tld{n}^2\,r^2+l^2\,p^2-3\,\tld{n}^4+l^2\,\tld{q}^2-2\,m\,l^2\,r}{l^2\,(r^2+\tld{n}^2)},
\end{equation}
and the gauge potential nonvanishing components are
\begin{equation}
    \tld{A}_t\,=\,\frac{\tld{n}\,\tld{p}+\tld{V}_e(r^2+\tld{n}^2)-\tld{q}\,r}{r^2+n^2}
\end{equation}
\begin{equation}
    \tld{A}_{\phi}\,=\,\frac{\tld{p}\,(\tld{n}^2-r^2)-2\,\tld{n}\,r\,\tld{q}}{l^2\,(r^2+\tld{n}^2)}\,x
\end{equation}
Thermodynamical quantities of this solution are
\bea
 && \tld{Q}_e\,={\sigma}\,(\tld{q}_e-2 \tld{n}\,\tld{\Phi}_m)=\tld{Q}^{\infty}_e-2 \tld{N}\,\tld{\Phi}_m, \hspace{0.5
in} \tld{\Phi}_e= \tld{V}_e \nonumber\\ &&\tld{{
Q}}_{m}\,=\,{\sigma}\,(\,\tld{p}\,+\,2\,\tld{n}\,\tld{\Phi}_e\,)=\tld{Q}^{\infty}_m+2 \tld{N}\,\tld{\Phi}_e,,\hspace{0.5
in}\tld{\Phi}_m= {\tld{p}+\tld{n}\,\tld{V}_e \over r_0}.
 \eea

Putting the metric in the ADM form
\begin{equation}
    ds^2\,=-\,N^2\,dt^2\,+\,h_{ij}\,(\,dx^i+\beta^i\,dt)\,(dx^j+\beta^j\,dt)
\end{equation}
where,
\begin{equation*}
    N^2\,=\,\frac{f(r)\,(r^2+\tld{n}^2)}{r^2+\tld{n}^2-4\,x^2\,\tld{n}^2\,f(r)\,l^{-2}},
\hspace{0.6 in}
    \beta^{\phi}\,=\,-\,\frac{2}{l} \left(
    \frac{\,\tld{n}\,x\,f(r)}{{r^2+\tld{n}^2-4\,x^2\,n^2\,f(r)l^{-2}}}\right),
\end{equation*}
where, $N$ is the lapse function and $\beta$ is the shift vector.

To do $3+1$ splitting, we start with a time-flow vector $t^\mu$,
defined as $t^\mu \nabla_\mu t\,=\,1$ or equivalently we have,
$t^\mu\,=\,\delta^{\mu}_0$. Also the spatial metric can be
constructed as follows,
\begin{equation}
    h^{\mu \nu}=g^{\mu \nu}+ n^{\mu}n^{\nu}
\end{equation}
where,
\begin{equation}
    n^{\mu}=\frac{1}{N}\,(\,t^{\mu}-\beta^{\mu}).
\end{equation}

In ADM $3+1$ split, we have a spatial hyper-surface $\Sigma$ with a
unit normal vector $n_{\mu}$, the metric on $\Sigma$ is $h^{\mu
\nu}$. In Hamiltonian formalism our dynamical fields are $h_{ab}$
and $A_a$ (after dropping tilde) where, $a,b=1,2,3$. the momenta of
the field variables are $\Pi_G^{a b}$ for the gravitational field
and $\pi^{a}$ for the electromagnetic field. where,
\begin{equation}
\Pi_G^{a b}=\frac{\partial L}{\partial \Dot{h}_{ab}}=-\frac{\sqrt{h}}{2 \kappa}\,(\,K^{a b}-h^{ab}\,K\,),
\end{equation}
\begin{equation}
 \Pi_{EM}^{a}=\frac{\partial L}{\partial \Dot{A}_{a}}=\frac{\sqrt{h}}{2 \kappa}\,(\,F^{\mu a}n_{\mu}\,)= \frac{\sqrt{h}}{2 \kappa}\,E^a,
\end{equation}
where, $K^{ab}$ is the extrinsic curvature of $h_{ab}$ and $K$ is its trace.

We have two constraints $C_\mu$, and $C$, where $C$ is Gaussian constraint or $D_a E^a=0$ while $C_\mu$ is the general relativity constraints determined by Einstein field equations, then one gets
\begin{equation}
    H=\int_{\Sigma } \xi^{\mu} C_{\mu}+\xi^{\mu} A_{\mu} C
\end{equation}
where, $\xi^{\mu}$ is the time evolution vector or time-like vector that vanishes at the horizon.
and,
\begin{equation} \label{Gaussconst}
    C=-\frac{\sqrt{h}}{4}\,D_a\,E^a
\end{equation}
\bea \label{GRconst1}
    C_{0}&&=-2\sqrt{h}\,(G_{\mu\nu}+\Lambda \, g_{\mu\nu}-2\,T_{\mu \nu})\,n^\mu n^\nu \nonumber\\
    &&=-\frac{\sqrt{h}}{4} R^{(3)}-{4 \sqrt{h}}(\Pi_G^{ab}\Pi^G_{ab}-{\Pi_G^2/2})+2\sqrt{h}\Lambda-8\frac{\pi^a \pi_a}{\sqrt{h}}-\frac{1}{ 4} \sqrt{h} F_{ab} F^{ab}\nonumber\\
\eea
\be \label{GRconst2}
    C_{a}=-2\,\sqrt{h}\,(G_{a\nu}+\Lambda \, g_{a\nu}-2\,T_{a \nu}) n^\nu = -2\,{\sqrt{h}}\,h_{ab}\, D_c({\Pi^{bc}\over \sqrt{h}})+4\,F_{ab}\,\pi^b.
\ee
The vanishing of the Hamiltonian leads to a relation between the mass and the other thermodynamic parameters \cite{horowitz}. Upon varying the mass with respect to electromagnetic quantities one gets,
\begin{equation}
   \delta M_{EM}= -\int dS_b \left[\xi^\mu A_\mu \delta E^b+(N F^{a b}-2\,E^{[a} \beta^{\,b]})\,\delta A_a
   \right],
\end{equation}
where $\xi$ is the time-like $\partial_t$. This is very similar to
the variation in \cite{horowitz}, but in four dimensions.
Calculating this expression and taking into account that for every
term we have three contributions coming from the boundary surfaces,
the constant-r surface $B_r$, and the two annuli $B_{x_{\pm}}$ one gets
the following result
\begin{equation}\label{deltaHem}
    \delta M_{EM}= \,\tld{V}_e\,d\,\tld{Q}^{\infty}_e+\, \tld{\Phi}_m\,d\,\tld{Q}_m
\end{equation}
which reproduces the electrodynamic part of first law in a more
formal manner. This confirms the previous thermodynamic
results.

\section{Conclusion}
I this work we present a new treatment for studying thermodynamics
and the first law of topological neutral and dyonic
Taub-Bolt/NUT-AdS solutions in four dimensions. This treatment is
based on introducing a new charge $N$ which is directly
related to the NUT charge, $n$. In our treatment, $N$ can vary
independently of the horizon radius $r_0$, as a result of the
absence of Misner string in the Euclidean Taub-Bolt/NUT solution.
Upon identifying one of the coordinates the spatial boundary of
these cases, the spatial boundary is a cylinder-like surface at
large radial distance $r$ with three distinguished surfaces, two at
constant-x and one at constant-r. Although the two constant-x
surfaces, or the annulus-like surfaces, do not receive fluxes of
conserved quantities in the $n=0$ case. The $n \neq 0$ case is
different and one can show that they bring additional contributions
to the mass, electric and magnetic charges, which are given as
$Q_e\sim q-2n\Phi_m$, $Q_m \sim p+2n\Phi_e$ and $\mathfrak{M} \sim
m-2n\Phi_n$. The calculated thermodynamic quantities obey the first
law of thermodynamics and the entropy is the area of the horizon.

Using $N=\sigma\,n$ as a new charge we were able to show that the first law in the neutral and dyonic
cases are satisfied using the quantities $Q_e$, $Q_m $ and
$\mathfrak{M}$, the entropy $S$, as the area of the horizon, $N$ and
$\Phi_N$. Furthermore, these quantities do satisfy Smarr's formula
in the neutral and dyonic cases. One of the intriguing issues of the
dyonic case is that although the full magnetic charge $Q_m$
contributes to the thermodynamics, only part of the electric charge,
namely, $Q^{\infty}_e$ contributes to it. To make sure that we got
the correct first law we followed the work in \cite{horowitz} to
calculate the Hamiltonian and its variation. Using the Hamiltonian
variation we obtained a similar formula to that of \cite{horowitz},
but in four dimensions, which reproduced the first law obtained in
the previous sections. This reflects the consistency of our
thermodynamic results.

It would be interesting to extend this treatment to Taub-Bolt
solutions with Lorentzian signature and spherical or hyperbolic
horizons which is under investigation now and we are going to report
on it soon. Another issue worth investigating is the reason the first law is only satisfied with
$Q^{\infty}_e$ and $Q_m$. We hope
that we can investigate this issue in some future work.

\appendix
\section{counter-term calculation}
In the following we use the counter-term method to calculate the action for a neutral Taub-Bolt solution,($q=p=V =0$). This method is equivalent to choosing a AdS as a background space-time (instead of the Taub-NUT background). Through calculating the thermodynamical quantities of this solution one can see that the first law and Smarr's formula are satisfied and have the same form presented in the article. We start with the gravitational action,\\
 \be I =I_{EH}+I_s+I_{ct}, \ee
 where, $I_{EH}$ is Einstein-Hilbert action, $I_s$ is Gibbons-Hawking boundary term, and $I_{ct}$ is a surface counter term added to cancel the divergences \cite{bala99,surf99}. The action can be written explicitly as,
\begin{equation}
 I = \frac{\sigma \beta}{2} \, \bigg[m+\frac{r_0\,(3n^2-r_0^2)}{l^2}\bigg]
\end{equation}
where $\beta$ is the temperature reciprocal, namely,
\begin{equation}
    T =\frac{1}{\beta} = \frac{3\,(r_0^2-n^2)}{4\pi\, l^2r_0}
\end{equation}
and,
\begin{equation}
 m = \frac{1}{2\,l^2\,r_0}\,(r_0^4-6\,n^2r_0^2-3\,n^4)
\end{equation}
Knowing that,
\begin{equation}
  G = \frac{I}{\beta} = H - TS
  \rightarrow H = \frac{I}{\beta} +TS
\end{equation}
where, $G$ is the Gibbs free energy, $H$ is the enthalpy. One can easily find that $H$ satisfies the following version of the first law,
\begin{equation}
 \delta H = T\delta S + V\delta P + \Phi_n \delta n
\end{equation}
 with $S$ is equal to a quarter of the horizon area, $V$ is the thermodynamic volume and here it takes the same form calculated in \cite{clifford14-1}, namely,
\begin{equation}
     V =\frac{\sigma}{3}(r_0^3-3n^2r_0)
\end{equation}
 and $\Phi_n$ is the nut potential which can be calculated through the action or Gibbs free energy as,
 \begin{equation}
     \Phi_n = \bigg(\frac{\partial G}{\partial n}\bigg) _{T,P} = -\frac{3\sigma n}{2\,l^2\,r_0}\,(r_0^2+n^2)
 \end{equation}
 Combining the above quantities, one can write Smarr`s formula of the mass $\mathfrak{M}$ as follows,
 \begin{equation}
     \mathfrak{M} = H -n\Phi_n = 2TS - 2PV
 \end{equation}
The variation of the Smarr`s formula satisfies the following version of the first law,
 \begin{equation}
   \delta \mathfrak{M} = T\delta S + V\delta P -n\,\delta \Phi_n
 \end{equation}
 which is in agreement with Smarr`s formulas and first laws in the case of Taub-NUT background. This calculation can be generalized to the charged case, $q\neq 0, p\neq 0, V\neq 0$, as well.

\section*{acknowledgments}
A. W. would like to thank Neil Lambert and Mohammad Alfiky for several
interesting discussion which helped us during the writing of this
work. This work is supported by Research Support Grant
(RSG1-2018-FY18-FY19) from the American University in Cairo; id:10.13039/501100009229, and
partially supported by the Egyptian Ministry of Scientific Research
project No. 24-2-12; id:10.13039/501100003007.

\end{document}